\documentclass[twocolumn,english]{article}
\usepackage[T1]{fontenc}
\usepackage{color}
\usepackage{array}
\usepackage{booktabs}
\usepackage{graphicx}
\usepackage{amssymb}

\makeatletter

\providecommand{\tabularnewline}{\\}

\usepackage{a4}
\input{epsfig.sty}
\def\fnum@table{\tablename~{\bf\thetable}}
\def\fnum@figure{\figurename~{\bf\thefigure}}
\def\tablename{\footnotesize{\bf Table}}
\def\figurename{\footnotesize{\bf Figure}}

\def\be{\begin{equation}}
\def\ee{\end{equation}}

\makeatother

\usepackage{babel}

\begin{document}

\title{\textbf{On the model-dependence of the relation between minimum-bias
and inelastic proton-proton cross sections}}

\author{\textbf{S. Ostapchenko}\\
\textit{\small NTNU, Institutt}{\small{} }\textit{\textcolor{black}{\small for
fysikk}}\textit{\small , 7491 Trondheim, Norway}\\
\textit{\small D.V. Skobeltsyn Institute of Nuclear Physics, Moscow
State University, 119992 Moscow, Russia}\textit{ }\\
}

\maketitle
\begin{center}
\textbf{\large Abstract}
\par\end{center}{\large \par}

The model-dependence of the relation between the inelastic and various
minimum-bias proton-proton cross sections is analyzed, paying a special
attention to the sensitivity of minimum-bias triggers to diffractive
collisions. Concentrating on the trigger selections of the ATLAS experiment,
the measured cross sections are compared to predictions of a number
of hadronic Monte Carlo models used in the cosmic ray field. It is
demonstrated that the ATLAS results are able to discriminate between
different models and between certain theoretical approaches for soft
multi-particle production. On the other hand, the strong model-dependence
of the selection efficiency of the minimum-bias triggers prevents
one from inferring high mass diffraction rate from the discussed data.
Moreover, the measured cross sections prove to be insensitive to the
production of low mass diffractive states in proton-proton collisions.
Consequently, a reliable determination of the total inelastic cross
section requires forward proton tracking by a dedicated experiment.

\section{Introduction\label{intro.sec} }

The knowledge of the inelastic proton-proton cross section and of
its repartition into the non-diffractive one and into partial cross
sections for various diffractive processes is of considerable importance
for understanding the dynamics of strong interactions. Additionally,
it is involved into determinations of collider luminosities and into
normalizations of measured particle spectra. On the other hand, a
proper understanding of the energy-dependence of the inelastic and
diffractive $pp$ cross sections is of vital importance for experimental
studies of high energy cosmic rays. Due to an extremely low incoming
flux of such particles, their properties are inferred from measured
characteristics of nuclear-electromagnetic cascades - extensive air
showers induced by them in the air. In turn, the longitudinal development
of air showers depends strongly on the magnitude of $\sigma_{pp}^{{\rm inel}}$
and on the relative rate of diffractive interactions \cite{ulr10}.

It has been proposed recently \cite{kmr09c} that a study of minimum-bias
cross sections at the Large Hadron Collider (LHC) with various combinations
of triggering detectors could be a powerful instrument for discriminating
between theoretical approaches to hadronic multiple production and
may allow one to infer the rate of diffraction at LHC energies. Presently,
such a study in underway by the ATLAS, CMS, and ALICE Collaborations
\cite{nav10} and the first results for the observed minimum-bias
cross sections have been reported by ATLAS \cite{aad11}.

The purpose of the present work is twofold. First, we investigate
model-dependence of the relation between various minimum-bias cross
sections and $\sigma_{pp}^{{\rm inel}}$, in particular, concerning
the contributions of low and high mass diffraction dissociation. Secondly,
we check if the ATLAS data are able to discriminate between different
models of hadronic interactions, in particular, between the ones used
to treat cosmic ray interactions in the atmosphere.

\section{Model approaches\label{sec:Model-approaches}}

General inelastic hadron-hadron collisions receive large 
contributions from soft processes and cannot be treated within the
perturbative QCD framework. This is especially so concerning the inelastic
diffraction which is often described in hadronic Monte Carlo (MC)
generators in a purely phenomenological way: based on empirical parametrizations
for the corresponding partial cross sections and assuming a simple
$dM_{X}^{2}/M_{X}^{2}$ distribution for the mass squared of diffractive
states produced \cite{schu94}. The only, though still phenomenological,
approach which offers a microscopic treatment of general soft and,
in particular, diffractive collisions of hadrons is provided by   
Gribov's Reggeon Field Theory (RFT) \cite{gri68}. In the RFT framework,
hadron-hadron collisions are described as multiple scattering processes,
with ``elementary'' scattering contributions being treated as Pomeron
exchanges. One usually employs separate treatments for low ($M_{X}^{2}\lesssim10$
GeV$^{2}$) and high mass diffractive excitations. The former can
be conveniently described using the Good-Walker-like multi-channel
approach \cite{goo60,kai79}: representing the interacting hadrons
by superpositions of a number of elastic scattering eigenstates characterized
by different interaction strengths and, generally, different transverse
profiles. Assuming eikonal vertices for Pomeron-hadron coupling, one
obtains simple expressions for partial cross sections of various low
mass diffraction processes. On the other hand, high mass diffraction
is related to the so-called enhanced diagrams which involve Pomeron-Pomeron
interactions \cite{kan73}, with the triple-Pomeron
coupling as the key parameter of the scheme. Importantly, at higher
energies enhanced graphs of more and more complicated topologies contribute
to elastic scattering amplitude and to partial cross sections of various,
notably diffractive, final states. Hence, meaningful results can only
be obtained after a full resummation of all significant enhanced contributions,
to all orders with respect to the triple-Pomeron coupling \cite{ost10}.

\begin{figure}[t]
\begin{centering}
\includegraphics[width=7cm,height=6cm]{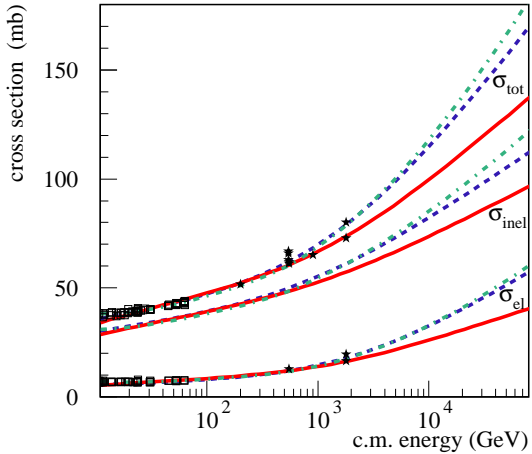}
\par\end{centering}

\caption{Model predictions for total, elastic, and inelastic proton-proton
cross sections: QGSJET-II-4 - solid, QGSJET-II-3 - dashed, and SIBYLL
- dot-dashed. The compilation of data is from Ref.~\cite{pdg10}.\label{fig:sigtot}}

\end{figure}

In this work, we analyze model-dependence of the relation between
$\sigma_{pp}^{{\rm inel}}$ and various minimum-bias proton-proton
cross sections, paying a special attention to the sensitivity of minimum-bias
triggers to diffractive collisions. We concentrate on the trigger
selections of the ATLAS experiment \cite{aad11} and employ in this
study hadronic MC generators used in the cosmic ray field. In particular,
we use the most recent version of the QGSJET-II model (QGSJET-II-04)
\cite{ost11} which is the only MC generator based on the full all-order
resummation of enhanced Pomeron graphs and which thus provides the
theoretically most advanced treatment of the physics relevant to the
present study.\footnote{Alternative approaches to the resummation of enhanced
Pomeron diagrams have been proposed in Refs.\
\cite{kmr08,kmr09b,kmr11,got08,got11,kol11}.}
 The corresponding results will be compared to the ones
of the SIBYLL model \cite{fle94,ahn09} which describes inelastic
diffraction similarly to the PYTHIA generator \cite{schu94,sjo06}:
$dM_{X}^{2}/M_{X}^{2}$ distribution is used for diffractive mass
squared and no special treatment for low mass diffraction is employed.
In addition, we use the previous version of QGSJET-II (QGSJET-II-03)
\cite{ost06b} which was based on the resummation of ``net''-like
enhanced Pomeron graphs while neglecting Pomeron ``loop'' contributions
\cite{ost06}. As a consequence, there is a factor of two difference
in the value of the triple-Pomeron coupling between the two model
versions, which projects itself in the corresponding difference for
the predicted single high mass diffraction cross section. On the other
hand, the inclusion of Pomeron loops significantly enhances the rate
of double high mass diffraction in QGSJET-II-04 compared to QGSJET-II-03.
While both model versions employ a 2-component multi-channel approach
for describing low mass diffraction, QGSJET-II-04 differs from the
previous version by using different transverse profiles for Pomeron
emission vertices by different elastic scattering eigenstates \cite{ost11}.%
\footnote{The choice of different profiles for the eigenstates was necessary
to get an agreement with measured differential elastic cross section
for hadron-proton collisions. On the other hand, such an agreement
could be obtained using the same transverse profile for all the eigenstates
but choosing a more sophisticated profile shape \cite{kmr09b}.%
} This had the consequence of a significantly larger low mass diffraction
rate at very high energies in QGSJET-II-04. The three models considered
differ also in their predictions for the high energy behavior of total
and inelastic proton-proton cross sections, as shown in Fig.~\ref{fig:sigtot},
 which is partly related to model calibrations to the CDF \cite{abe94}
or E710 \cite{amo92} results on $\sigma_{pp}^{{\rm tot}}$ at the
Tevatron. A comparison of the three models' results with various accelerator
data, from  fixed target energies up to LHC, may be found in
 Refs.\ \cite{ost11,ahn09,ost06b,den11}.

In Table~\ref{Flo:sigmas} %
\begin{table*}[t]
\centering{}\begin{tabular}{lccccc}
\hline
\hline 
 & {\small $\sigma_{{\rm inel}}$} & {\small $\sigma_{{\rm ND}}$} 
 & {\small $\sigma_{{\rm SD}}^{{\rm HM}}+\sigma_{{\rm DD}}^{{\rm LHM}}$}
  & {\small $\sigma_{{\rm DD}}^{{\rm HM}}$} 
   & {\small $\sigma_{{\rm SD}}^{{\rm LM}}+\sigma_{{\rm DD}}^{{\rm LM}}$}
   \tabularnewline
\midrule 
{\small QGSJET~II-04} & {\small 69.7} &  {\small 49.6} &  {\small 5.7} &   {\small 7.3} &   {\small 7.1}\tabularnewline
{\small QGSJET~II-03} & {\small 77.5} &  {\small 57.4} &  {\small 11.4} &   {\small 5.4} &   {\small 3.3}\tabularnewline
{\small SIBYLL} & {\small 79.6} &  {\small 65.7} &  {\small 12.2} &   {\small 1.7} &   {\small 0}\tabularnewline
{\small PYTHIA} & {\small 71.5} &  {\small 48.5} &  {\small 13.7} &   {\small 9.3} &   {\small 0}\tabularnewline
\hline
\hline
\end{tabular}\caption{Model predictions for $\sigma_{pp}^{{\rm inel}}$ and for partial
inelastic proton-proton cross sections (in mb) at $\sqrt{s}=7$ TeV.}
\label{Flo:sigmas}
\end{table*}
we compile the predictions of the above-discussed models for $\sigma_{pp}^{{\rm inel}}$
at $\sqrt{s}=7$ TeV as well as the respective partial cross sections
for non-diffractive $\sigma_{{\rm ND}}$, single high mass diffractive
$\sigma_{{\rm SD}}^{{\rm HM}}$ (combined with double diffractive
cross section $\sigma_{{\rm DD}}^{{\rm LHM}}$ corresponding to a
high mass diffraction of one proton and a low mass excitation of the
other one), double high mass diffractive $\sigma_{{\rm DD}}^{{\rm HM}}$,
and single $\sigma_{{\rm SD}}^{{\rm LM}}$ and double $\sigma_{{\rm DD}}^{{\rm LM}}$
low mass diffractive collisions.%
\footnote{In Table~\ref{Flo:sigmas} we use \textsl{theoretical} definitions
for low and high mass diffraction cross sections, which do not depend
on an experimental trigger. For comparison, in \cite{ost11} the diffractive
event classification of the CDF experiment was applied to compare
with the corresponding data.%
} For comparison we add also the corresponding PYTHIA results taken
from Ref.~\cite{aad11}. The considered subclasses of inelastic collisions
differ in their efficiencies to generate a signal in scintillation
detectors used in minimum-bias trigger selections (MBTS) of various
LHC experiments. In non-diffractive events, all the kinematically
accessible rapidity interval is covered by secondaries, hence, the
probability to have a charged hadron inside a detector should approach
100\% in that case.%
\footnote{In QGSJET-II, there is a small (sub-mb) contribution of central diffraction
(double Pomeron exchange) which we did not subtract from $\sigma_{{\rm ND}}$.%
} In turn, high mass diffractive collisions contain large rapidity
gaps not covered by secondary particles. Hence, a noticeable part
of such collisions will be missed by the MBTS detectors - when the
respective rapidity coverage of the detectors is fully inside the
rapidity gaps. Finally, low mass diffractive interactions produce
narrow bunches of secondaries in forward and/or backward hemisphere,
such that almost the whole rapidity range, except its edges, is free
of secondaries. Such events are likely to be missed by the triggers,%
\footnote{In the QGSJET-II model, the low mass diffraction cross sections calculated
in the multi-channel approach are assumed to correspond to diffractive
final states described by the Pomeron-Pomeron-Reggeon ($\mathbb{PPR}$)
asymptotics, with an approximate $dM_{X}^{2}/M_{X}^{3}$ diffractive
mass distribution. Hence, one obtains some weak sensitivity of the
MBTS detectors to the tail of the $M_{X}$ distribution, as demonstrated
in the following. Alternatively, one may assume that these cross sections
correspond to a number of discrete low mass resonance states \cite{kmr09b},
in which case such events would be missed completely by the triggers.%
} which was in particular the reason for combining $\sigma_{{\rm SD}}^{{\rm HM}}$
and $\sigma_{{\rm DD}}^{{\rm LHM}}$ together.

\begin{figure}[t]
\begin{centering}
\includegraphics[width=7cm,height=6cm]{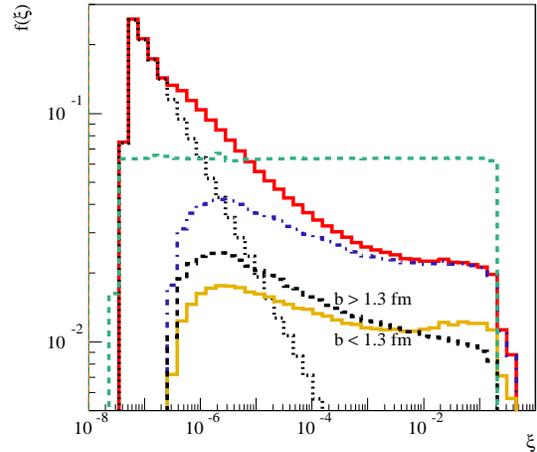}
\par\end{centering}

\caption{$f_{{\rm SD}}(\xi)\equiv\frac{\xi}{\sigma_{{\rm SD}}}\,\frac{d\sigma_{{\rm SD}}}{d\xi}$
for single diffractive $pp$ collisions at $\sqrt{s}=7$ TeV as calculated
using QGSJET-II-4 (red solid) and SIBYLL (green dashed). Partial contributions
to $f_{{\rm SD}}(\xi)$ from low and high mass diffraction in QGSJET-II-4
are shown as black doted and blue dash-dotted lines respectively;
yellow solid and black dashed lines are used for contributions of
high mass diffraction at $b<1.3$ fm and $b>1.3$ fm.\label{fig:M_X}}

\end{figure}

Importantly,  the considered models differ not only in the
predicted diffraction cross sections but also concerning mass distributions
of diffractive states. This is illustrated in Fig.~\ref{fig:M_X},%
 where the distribution for $\xi=M_{X}^{2}/s$, $f_{{\rm SD}}(\xi)\equiv\frac{\xi}{\sigma_{{\rm SD}}}\,\frac{d\sigma_{{\rm SD}}}{d\xi}$,
is plotted for single diffractive $pp$ collisions at $\sqrt{s}=7$
TeV, as calculated using QGSJET-II-04 and SIBYLL. In contrast to the
flat $\xi$-distribution of the SIBYLL model, which corresponds to
the assumed $dM_{X}^{2}/M_{X}^{2}$ distribution for diffractive masses,
a more complicated functional shape is predicted by QGSJET-II-04.
The most striking difference is the sharp peak at $\xi\sim{\rm few}\:{\rm GeV}^{2}/s$
corresponding to the production of low mass diffractive states. Additionally,
the distribution for high mass diffraction changes from the $1/M_{X}^{2}$
behavior at very large $M_{X}^{2}$ to a steeper decreasing function
at smaller $M_{X}^{2}$, which reflects the impact parameter $b$
dependence of absorptive corrections to diffractive scattering. As
noted in \cite{ost10}, strong absorptive corrections at small $b$
result in the approximate $f(M_{X}^{2})\sim1/M_{X}^{2}$ distribution
for diffractive masses. On the other hand, in  very peripheral collisions
absorptive corrections become weak and high mass diffraction is governed
by the triple-Pomeron contribution with $f(M_{X}^{2})\sim(M_{X}^{2})^{-\alpha_{\mathbb{P}}}$
(in QGSJET-II-04 the soft Pomeron intercept $\alpha_{\mathbb{P}}=1.16$
\cite{ost11}). As a cross-check, one may compare two sub-samples
of the simulated high mass diffraction collisions, with approximately
equal event numbers: more peripheral collisions with $b>1.3$ fm and
more central ones ($b<1.3$ fm). As one can see in Fig.~\ref{fig:M_X},
the latter case leads indeed to a flatter distribution for the diffractive
mass. 

Similarly, there are large differences between QGSJET-II and SIBYLL
in the treatment of double high mass diffraction. As demonstrated
in \cite{ost10}, already at the lowest order with respect to the
triple-Pomeron coupling the corresponding picture is considerably
more complicated than usually assumed in literature. Double diffractive
final states may result from a collision which contains a single inelastic
rescattering process characterized by the desirable structure of the
final state (central rapidity gap) or from a collision with two inelastic
rescatterings: one corresponding to single high mass diffraction of
the projectile and the other - of the target, and with the corresponding
rapidity gaps overlapping in the central region. Moreover, there exists
a non-trivial interference between the two contributions \cite{ost10}.

\section{Results for minimum-bias cross sections\label{sec:triggers}}

The models discussed in the previous section have been used to generate
hadronic final states corresponding to non-diffractive and various
diffractive proton-proton collisions and to investigate selection
efficiencies $\varepsilon$ of the corresponding final states by minimum-bias
triggers of the ATLAS experiment. Here we restrict ourselves with
the MBTS\_AND and MBTS\_OR triggers which require at least one charged
particle detected respectively at both positive ($\eta_{1}<\eta<\eta_{2}$)
and negative ($-\eta_{2}<\eta<-\eta_{1}$) pseudorapidity intervals
or in either of the two $\eta$-ranges, with $\eta_{1}=2.09$ and
$\eta_{2}=3.84$ \cite{aad11}. In addition, we consider the ChPart
trigger selection which combines the MBTS\_OR condition with the requirement
of at least one charged hadron of transverse momentum $p_{t}>0.5$
GeV being detected in the $|\eta|<0.8$ range. When estimating the
corresponding selection efficiencies we assume 100\% detection probability
for charged hadrons in the respective pseudorapidity intervals, thus
neglecting potential loss of events due to less than 100\% particle
tracking efficiency. Though more accurate treatment, with the tracking
efficiency properly taken into account, may result in somewhat lower
trigger rates for inelastic processes, we do not expect the difference
to be large because of the relatively wide pseudorapidity intervals
involved and high particle densities produced.%
\footnote{The possible exception is the ChPart selection which, however, has
been corrected for the particle tracking efficiency of the ATLAS experiment
in Ref.~\cite{aad11}.%
}

The obtained model results for selection efficiencies of various inelastic
processes, compiled in Table~\ref{Flo:effic-Q-04},%
\begin{table*}[tbh]
\centering{}\begin{tabular}{lccccccccccc}
\hline 
\hline 
 &\multicolumn{4}{c}{{\small QGSJET-II-04}}     &\multicolumn{4}{c}{{\small QGSJET-II-03}} 
  &\multicolumn{3}{c}{{\small SIBYLL}}  \\
 & {\small $\varepsilon_{{\rm ND}}$} & {\small $\varepsilon_{{\rm SD}}^{{\rm HM}}$}
  & {\small $\varepsilon_{{\rm DD}}^{{\rm HM}}$}  & {\small $\varepsilon_{{\rm LMD}}$}
   & {\small $\varepsilon_{{\rm ND}}$} & {\small $\varepsilon_{{\rm SD}}^{{\rm HM}}$} 
   & {\small $\varepsilon_{{\rm DD}}^{{\rm HM}}$}  & {\small $\varepsilon_{{\rm LMD}}$}
    & {\small $\varepsilon_{{\rm ND}}$} & {\small $\varepsilon_{{\rm SD}}^{{\rm HM}}$} 
    & {\small $\varepsilon_{{\rm DD}}^{{\rm HM}}$} \tabularnewline
\hline 
{\small MBTS\_AND} & {\small 95} & {\small 25} & {\small 76} & {\small 0.3} & {\small 95} & {\small 29} & {\small 82} & {\small 0.2} & {\small 98} & {\small 28} & {\small 59}\tabularnewline
{\small MBTS\_OR} & {\small 100} & {\small 64} & {\small 96} & {\small 7.1} & {\small 100} & {\small 63} & {\small 97} & {\small 6.4} & {\small 100} & {\small 62} & {\small 86}\tabularnewline
{\small ChPart} & {\small 87} & {\small 24} & {\small 56} & {\small 0.5} & {\small 89} & {\small 29} & {\small 66} & {\small 0.3} & {\small 89} & {\small 25} & {\small 44}\tabularnewline
\hline 
\hline
\end{tabular}\caption{Model results for selection efficiencies (in \%) of the ATLAS minimum-bias
triggers to non-diffractive ($\varepsilon_{{\rm ND}}$), single high
mass diffractive ($\varepsilon_{{\rm SD}}^{{\rm HM}}$), double high
mass diffractive ($\varepsilon_{{\rm DD}}^{{\rm HM}}$), and low mass
diffractive ($\varepsilon_{{\rm LMD}}$) proton-proton collisions
at $\sqrt{s}=7$ TeV.}
\label{Flo:effic-Q-04}
\end{table*}
 confirm qualitative expectations of Section~\ref{sec:Model-approaches},
demonstrating in particular that the ATLAS trigger selections have
high efficiency for triggering non-diffractive interactions while
being almost blind to low mass diffraction. On the other hand, we
observe large differences between QGSJET-II and SIBYLL for $\varepsilon_{{\rm DD}}^{{\rm HM}}$,
which is related to the treatment of double high mass diffraction
in the two models. The contribution to $\sigma_{{\rm DD}}^{{\rm HM}}$
from the superposition of two simultaneous (projectile and target)
single diffraction processes results in much narrower rapidity gap
sizes (hence, in a smaller probability to miss the trigger) compared
to the case of simple $dM_{X}^{2}/M_{X}^{2}$ distributions for the
two diffractive masses, used in SIBYLL. It is noteworthy that $\varepsilon_{{\rm DD}}^{{\rm HM}}$
in QGSJET-II-03 is higher than in QGSJET-II-04 because Pomeron loops
are neglected in the former model, hence, the above-discussed mechanism
is the only one relevant for double high mass diffraction in that
case. As anticipated in \cite{kmr09c}, the MBTS\_AND and MBTS\_OR
triggers differ significantly in their sensitivity to single high
mass diffraction: the ratio of the corresponding values for $\varepsilon_{{\rm SD}}^{{\rm HM}}$
is related to the ratio of the numbers of events with $\xi>\ln s/2+\eta_{1}$
and $\xi>\ln s/2-\eta_{2}$. This explains also the larger difference
between the respective values of $\varepsilon_{{\rm SD}}^{{\rm HM}}$
for the two triggers in case of QGSJET-II-04 compared to SIBYLL - as
in the former model single high mass diffractive events have a smaller
probability to fall in the interval $\xi>\ln s/2+\eta_{1}$ (c.f.~blue
dot-dashed and green dashed lines in Fig.~\ref{fig:M_X}). However,
the values of $\varepsilon_{{\rm SD}}^{{\rm HM}}$ for QGSJET-II-03
appear to be similar to the ones for SIBYLL, which indicates that
other effects, e.g.~the rapidity density of produced hadrons, impact
the results.

In Table~\ref{Flo:sigma-vis}, %
\begin{table*}[t]
\centering{}\begin{tabular}{lccccc}
\hline 
\hline 
 & {\small QGSJET-II-04} & {\small QGSJET-II-03} & {\small SIBYLL} & {\small PYTHIA} & {\small Exp.~\cite{aad11}}\tabularnewline
\hline 
{\small MBTS\_AND} & {\small 54.1} & {\small 62.3} & {\small 68.4} & {\small 58.4} & {\small 51.9$\pm$0.2}\tabularnewline
{\small MBTS\_OR} & {\small 60.8} & {\small 69.8} & {\small 74.7} & {\small 66.6} & {\small 58.7$\pm$0.2}\tabularnewline
{\small ChPart} & {\small 48.4} & {\small 57.7} & {\small 62.3} & {\small 45.7} & {\small 42.7$\pm$0.2}\tabularnewline
\hline 
\hline
\end{tabular}\caption{Model predictions for visible cross sections (in mb) in the ATLAS
minimum-bias detectors compared to the measured values \cite{aad11}
(the correlated 11\% uncertainty of the experimental results related
to the luminosity determination is not included in the quoted experimental
errors).}
\label{Flo:sigma-vis}
\end{table*}
we compare the model predictions for ``visible'' cross sections for
the various combinations of ATLAS minimum-bias triggers with experimental
results; the corresponding values for PYTHIA from Ref.~\cite{aad11}
are also added. It is easy to see that model extrapolations based
on the CDF measurement of $\sigma_{pp}^{{\rm tot}}$ at the Tevatron
are disfavored by the ATLAS results. Indeed, of the three models considered
only QGSJET-II-04 agrees with the data within the uncertainties related
to the luminosity determination. QGSJET-II-03 and SIBYLL exceed the
measured MBTS rates by about $2\sigma$ and $3\sigma$ respectively
while even stronger disagreement is observed for the ChPart event
selection. On the other hand, the experimental results demonstrate
the potential for discriminating between various theoretical models
for soft multi-particle production. As an example, one may consider
the approach of Refs.~\cite{kmr08,kmr09b}, which predicts
a much slower energy rise of the total and inelastic proton-proton
cross sections and a significantly higher $\sigma_{{\rm SD}}^{{\rm HM}}$
compared to the one of Refs.~\cite{ost10,ost08}, realized in QGSJET-II-04.
The mentioned differences are mostly due to specific assumptions on
Pomeron-Pomeron interaction vertices, made in \cite{kmr08,kmr09b},
such that the scheme approaches the so-called ``critical'' Pomeron
description in the ``dense'' limit of high energies and small impact
parameters, as discussed in more detail in \cite{ost10,kmr11}. In
turn, this results in much smaller visible cross sections for the
ATLAS minimum-bias triggers \cite{kmr09c}, which appear to be some
10 mb below the measured values. 

One may doubt if the results of Ref.~\cite{kmr09c} may be modified
by hadronization effects which have not been included in the corresponding
analysis. We check such a possibility with QGSJET-II-04 by comparing
the MBTS efficiencies for single and double high mass diffractive
events obtained in two ways: based on charged particle tracking (as
summarized in Table~\ref{Flo:effic-Q-04}) or using the \textsl{theoretical
}rapidity gap structure of individual events. In the latter case,
we assume that an event is triggered if the rapidity coverage of the
respective detectors is at least partly spanned by a cut Pomeron (which
corresponds to an elementary ``piece'' of particle production in the
model) produced when modeling the configuration of the interaction
\cite{ost11}. Alternatively, an event is not triggered if the detectors
appear to be fully inside the \textsl{theoretical }rapidity gaps (defined
by the cut Pomeron structure of the event). As is easy to see from
Table~\ref{Flo:effic-theor}, %
\begin{table}[tbh]
\centering{}{\small }\begin{tabular}{>{\raggedright}p{19mm}>{\centering}p{4mm}>{\centering}p{4mm}>{\centering}p{4mm}>{\centering}p{4mm}>{\centering}p{12mm}}
\hline 
\hline 
 & {\small $\varepsilon_{{\rm ND}}$} & {\small $\varepsilon_{{\rm SD}}^{{\rm HM}}$} & {\small $\varepsilon_{{\rm DD}}^{{\rm HM}}$ } & {\small $\varepsilon_{{\rm LMD}}$} & {\small $\sigma_{{\rm obs}}$}\tabularnewline
\hline 
{\small MBTS\_AND} & {\small 100} & {\small 20} & {\small 54} & {\small 0} & {\small 54.9 mb}\tabularnewline
{\small MBTS\_OR} & {\small 100} & {\small 66} & {\small 91} & {\small 0} & {\small 60.3 mb}\tabularnewline
\hline 
\hline
\end{tabular}\caption{Selection efficiencies (in \%) of various partial inelastic $pp$
cross sections at $\sqrt{s}=7$ TeV by the ATLAS minimum-bias triggers
and the resulting visible cross sections, as obtained from the theoretical
rapidity gap structure of final states in QGSJET-II-04.}
\label{Flo:effic-theor}
\end{table}
the corresponding theoretical efficiencies for the MBTS\_OR selection
coincide with the full MC results (Table~\ref{Flo:effic-Q-04}) within
few percents thereby confirming that the results of Ref.~\cite{kmr09c}
are indeed disfavored by the data. On the other hand, the selection
efficiencies for the MBTS\_AND trigger appear to be strongly modified
by hadronization effects, which indicates that the latter change noticeably
the rapidity gap structure of individual events. Hence, a complete
MC treatment is a necessary pre-requisite for inferring the high mass
diffraction rate from a set of minimum-bias cross sections.

Clearly, without a good handle on the low mass diffraction,
a determination of $\sigma_{pp}^{{\rm inel}}$
on the basis of measured minimum-bias cross sections will be strongly
model-dependent. On the other hand,
one may be tempted to use the ATLAS results to
 derive luminosity-independent estimations of
the relative contributions of single and double high mass diffraction,
$(\sigma_{{\rm SD}}^{{\rm HM}}+\sigma_{{\rm DD}}^{{\rm LHM}})/\sigma_{{\rm abs}}^{{\rm HM}}$,
$\sigma_{{\rm DD}}^{{\rm HM}}/\sigma_{{\rm abs}}^{{\rm HM}}$, with
$\sigma_{{\rm abs}}^{{\rm HM}}\equiv\sigma_{{\rm inel}}-\sigma_{{\rm SD}}^{{\rm LM}}-\sigma_{{\rm DD}}^{{\rm LM}}$.
Such an information would be extremely valuable since model predictions
for $\sigma_{{\rm SD}}^{{\rm HM}}$ and $\sigma_{{\rm DD}}^{{\rm HM}}$
are vastly different, as demonstrated in Table~\ref{Flo:sigmas}.
However, the corresponding equation system for the ATLAS trigger selections\begin{eqnarray*}
\sigma_{{\rm MBTS-AND}} & = & \varepsilon_{{\rm ND}}^{{\rm AND}}\sigma_{ND}+\varepsilon_{{\rm SD(HM)}}^{{\rm AND}}(\sigma_{{\rm SD}}^{{\rm HM}}\\
 &  & +\sigma_{{\rm DD}}^{{\rm LHM}})+\varepsilon_{{\rm DD(HM)}}^{{\rm AND}}\sigma_{{\rm DD}}^{{\rm HM}}\\
\sigma_{{\rm MBTS-OR}} & = & \varepsilon_{{\rm ND}}^{{\rm OR}}\sigma_{ND}+\varepsilon_{{\rm SD(HM)}}^{{\rm OR}}(\sigma_{{\rm SD}}^{{\rm HM}}\\
 &  & +\sigma_{{\rm DD}}^{{\rm LHM}})+\varepsilon_{{\rm DD(HM)}}^{{\rm OR}}\sigma_{{\rm DD}}^{{\rm HM}}\\
\sigma_{{\rm ChPart}} & = & \varepsilon_{{\rm ND}}^{{\rm ChPart}}\sigma_{ND}+\varepsilon_{{\rm SD(HM)}}^{{\rm ChPart}}(\sigma_{{\rm SD}}^{{\rm HM}}\\
 &  & +\sigma_{{\rm DD}}^{{\rm LHM}})+\varepsilon_{{\rm DD(HM)}}^{{\rm ChPart}}\sigma_{{\rm DD}}^{{\rm HM}}\end{eqnarray*}
is ill-defined, mainly because of the strong correlation between the
ChPart and MBTS triggers and due to the strong model-dependence of
the selection efficiency for double high mass diffraction. Alternatively,
one may combine non-diffractive and double high mass diffractive events
together and restrict oneself with the MBTS results only. Applying
a MC procedure to determine the selection efficiency for non-single-diffractive
events ($\sigma_{NSD}\equiv\sigma_{ND}+\sigma_{DD}^{{\rm HM}})$,
we obtain using QGSJET-II-04, QGSJET-II-03, and SIBYLL the results
listed in Table~\ref{Flo:effic-nsd}.%
\begin{table}[t]
\centering{}{\small
}\begin{tabular}{>{\raggedright}p{11mm}>{\centering}p{19mm}>{\centering}p{19mm}>{\centering}p{8mm}}
\hline 
\hline 
 &{\scriptsize QGSJET-II-04}&{\scriptsize QGSJET-II-03}&{\scriptsize SIBYLL}\tabularnewline
\hline 
{\scriptsize  MBTS\_AND} & {\small 92} & {\small 94} & {\small 97}\tabularnewline
{\scriptsize  MBTS\_OR} & {\small 99} & {\small 100} & {\small 100}\tabularnewline
\hline 
\hline
\end{tabular}\caption{Model results for the selection efficiency $\varepsilon_{{\rm NSD}}$
(in \%) of non-single diffractive events by the ATLAS minimum-bias
triggers.}
\label{Flo:effic-nsd}
\end{table}
The corresponding values of $(\sigma_{{\rm SD}}^{{\rm HM}}+\sigma_{{\rm DD}}^{{\rm LHM}})/\sigma_{{\rm abs}}^{{\rm HM}}$
and of partial cross sections are compiled in Table~\ref{Flo:sigma-ratio}.%
\begin{table}[t]
\centering{}{\small
}\begin{tabular}{>{\raggedright}p{15.1mm}>{\centering}p{18.1mm}>{\centering}p{18.1mm}>{\centering}p{8mm}}
\hline 
\hline 
 & {\scriptsize QGSJET-II-04} & {\scriptsize QGSJET-II-03} & {\scriptsize SIBYLL}\tabularnewline
\hline 
{\scriptsize $\frac{\sigma_{{\rm SD}}^{{\rm HM}}+\sigma_{{\rm DD}}^{{\rm LHM}}}{\sigma_{{\rm abs}}^{{\rm HM}}}$} & {\footnotesize0.12} & {\footnotesize0.19} & {\footnotesize0.24}\tabularnewline
{\scriptsize $\sigma_{{\rm SD}}^{{\rm HM}}+\sigma_{{\rm DD}}^{{\rm LHM}}$} & {\footnotesize7.4 mb} & {\footnotesize12 mb} & {\footnotesize16 mb}\tabularnewline
{\scriptsize $\sigma_{{\rm NSD}}$} & {\footnotesize54 mb} & {\footnotesize51 mb} & {\footnotesize49 mb}\tabularnewline
{\scriptsize $\sigma_{{\rm abs}}^{{\rm HM}}$} & {\footnotesize62 mb} & {\footnotesize63 mb} & {\footnotesize65 mb}\tabularnewline
\hline 
\hline
\end{tabular}\caption{Model-based estimations of the relative fraction of single high mass
diffraction and of partial inelastic cross sections.}
\label{Flo:sigma-ratio}
\end{table}
It is easy to see that the obtained relative rate of single high mass
diffraction is strongly model-dependent. Hence, a direct measurement
of high mass diffraction is required to discriminate between the various
model approaches.

\section{Outlook\label{sec:Outlook}}

In this work, we analyzed the model-dependence of the relation between
the inelastic and various minimum-bias proton-proton cross sections,
concentrating on the trigger selections of the ATLAS experiment and
comparing the measured visible cross sections at $\sqrt{s}=7$ TeV
to predictions of a number of hadronic Monte Carlo generators used
in the cosmic ray field. We demonstrated that the ATLAS results provide
serious constraints on hadronic interaction models and 
allow one to discriminate between certain theoretical approaches
to the treatment of soft multi-particle production. In particular,
 the minimum-bias trigger rates reported by ATLAS  disfavor model
extrapolations based on the CDF measurement of $\sigma_{pp}^{{\rm tot}}$
at the Tevatron.  This will have an important impact on the studies
of the nuclear composition of ultra-high energy cosmic rays with
fluorescence light detectors \cite{abr10,abb10}. 

On the other hand, the strong model-dependence of the sensitivity
of the ATLAS minimum-bias triggers to diffractive events, mainly due to the
differences in the predicted diffractive mass distributions, prevents
one from deducing the high mass diffraction rate from the measured
visible cross sections. Moreover, the MBTS detectors prove to be insensitive
to the contribution of low mass diffraction in proton-proton collisions.
Hence, a reliable determination of the total inelastic $pp$ cross
section does not seem feasible without forward proton tracking by
a dedicated experiment, like TOTEM \cite{ane08}.

\vspace{5mm}

\noindent{\bf \small Note added.} 

\vspace{3mm}

\noindent{\small After this paper was submitted to the journal,
a measurement of the  inelastic proton-proton cross section
for  $\xi > 5\times 10^{-6}$
has been reported by the ATLAS Collaboration \cite{aad11a}.
Extrapolation of the result to the full kinematic range was
found to be strongly model-dependent, primarily, due to strong
sensitivity to model predictions for the diffractive mass distribution
-- which was also the main source of model-dependence discussed in this work.}

\subsection*{Acknowledgments}

The author acknowledges useful discussions with David d'Enterria and
the  support  by the program Romforskning of Norsk Forskningsradet.

\end{document}